\documentclass[intlimits,twoside,a4paper]{article}

\usepackage[cp1251]{inputenc}
\usepackage{multirow}
\usepackage[eqsecnum]{cmpj3}

\usepackage{bm}
\newcommand{\Det}{\text{Det}}
\newcommand{\Tr}{\text{Tr}}

\issue{2021}{24}{2}{23301}
\doinumber{10.5488/CMP.24.23301}

\title[MC investigations of higher generation ideal dendrimers]%
{Monte Carlo computer investigations of higher generation ideal dendrimers%
}

\author[M. Jura, M. Bishop, B. Thrope, R. de Regt]{M.  Jura\orcid{0000-0002-3424-4172}\refaddr{label1}\thanks{Corresponding author: \email{matthew.jura@manhattan.edu}.},
        M. Bishop\orcid{0000-0003-1774-1238}\refaddr{label1}, B. Thrope\orcid{0000-0002-6307-1178}\refaddr{label1},
        R. de Regt\orcid{0000-0002-8760-0193}\refaddr{label2, label3}}
\addresses{
\addr{label1} Department of Mathematics, Manhattan College, Manhattan College Parkway, Riverdale, NY 10471, USA
\addr{label2} Applied Mathematics Research Centre, Coventry University, Coventry CV1 5FB, UK 
\addr{label3}  Doctoral College for the Statistical Physics of Complex Systems,
Leipzig-Lorraine-Lviv-Coventry(L4), D-04009 Leipzig, Germany 
}

\date{Received August 16, 2020, in final form February 5, 2021}
\begin{document}

\maketitle

\begin{abstract}
The properties of ideal tri-functional dendrimers with forty-five, ninety-three and one hundred and eighty-nine branches are investigated. Three methods are employed to calculate the mean-square radius of gyration, $g$-ratios, asphericity, shape parameters and form factor. These methods include a Kirchhoff matrix eigenvalue technique, the graph theory approach of Benhamou et al. (2004), and Monte Carlo simulations using a growth algorithm. A novel technique for counting  paths in the graph representation of the dendrimers is presented. All the methods are in excellent agreement with each other and with available theoretical predictions. Dendrimers become more symmetrical as the generation and the number of branches increase.
\printkeywords
%
\end{abstract}

\section{Introduction}

The architecture of a polymer is determined by how its units are linked together. The simplest polymer is a linear chain. Three or more linear chains can be grown from a single junction to form a star polymer. If the ends of each chain are then used in turn repeatedly as junctions for a further growth one obtains a dendrimer. These polymers have the shape of a branching tree and can be classified by their functionality (number of linear branches growing from a junction), the spacer size (the number of units in a linear branch), and the generation. A generation is defined by the number of rings of branches starting from a single central junction.

\begin{figure}[!t]
\centerline{\includegraphics[width=0.5\textwidth]{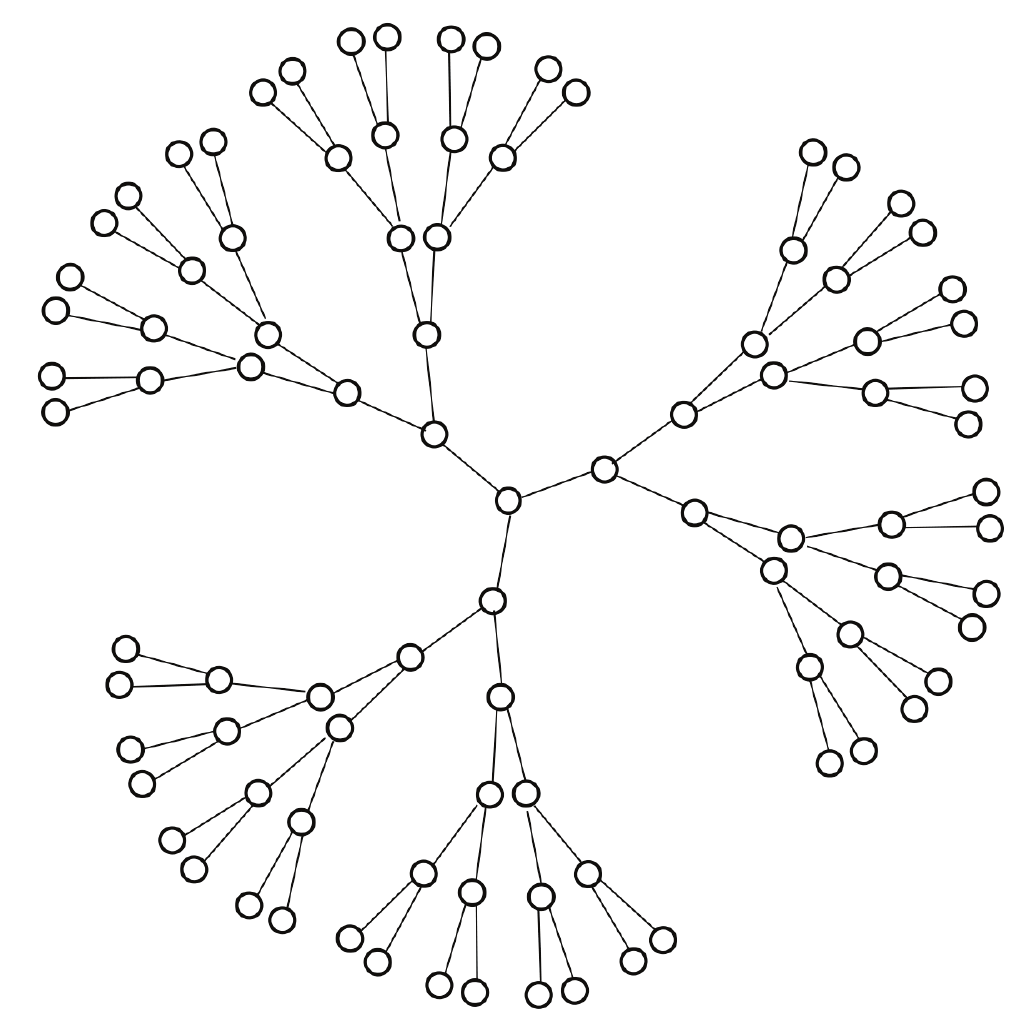}}
\caption{93-branch dendrimer (fourth generation).} 
\label{93dendrimer}
\end{figure}

Figure \ref{93dendrimer} illustrates a typical dendrimer. The diagram displays a 93-branch dendrimer. This is the fourth generation in its family. The circles represent the junctions. The first three branches attached to the central junction represent a star polymer. The next ring, containing six more branches, represents the first dendrimer generation, a molecule with nine branches. The second generation has twenty-one branches, the third, forty-five and the fourth, ninety-three. The fifth generation will have one hundred and eighty-nine branches so that its structure would include another ring of branches. Two more branches would extend from each circle, adding 96 branches to the structure for a total of 189. The reason for the 9-21-45-93-189 branch numbers is that the number being added doubles; 12 branches are added to 9 to get 21, 24 to 21 to get 45, etc. The number of units in each branch is uniform throughout the entire structure and can be any value. The total number of units, $N$, in each dendrimer is $N = f m+1$ where $f$ is the number of branches defined by the dendrimer generation and $m$ is the number of units in each branch. In our study, $N$ ranged from 901 to 4915.

 The mean-square radius of gyration \cite{F1}, $\langle S^2 \rangle$, is used to measure the total size of a dendrimer. The mean-square radius of gyration of an object composed of $ N$ identical units, around its center of mass, is
\begin{equation} \label{e:g}
   \langle S^2 \rangle = \frac {1}{N} \sum_i^N \left\langle \left( X_i-X_{CM}\right)^2 + \left(Y_i-Y_{CM}\right)^2 + \left(Z_i-Z_{CM}\right)^2\right\rangle,   
\end{equation}	
where $X_i, Y_i$ and $Z_i$ are the components of the position of the $i$-th unit and the center of mass is defined by
\begin{equation} \label{e:cm}
   X_{CM} = \frac {1}{N} \sum_i^N  X_i;\ \  Y_{CM} = \frac {1}{N} \sum_i^N  Y_i; \ \ 
   Z_{CM} = \frac {1}{N} \sum_i^N  Z_i. 
\end{equation}	
The braces $\langle \quad\rangle$ indicate an average over multiple dendrimer constructions.

Sheng et al. \cite{SJT} determined an exact equation for the radius of gyration of ideal tri-functional dendri\-mers:
\begin{equation} \label{e:exact}
 \frac {\langle S^2 \rangle N^2}{\langle l^2 \rangle} = 3m^3\left[ -1+6\cdot 2^{G1}+(3G1-5)2^{2G1}\right]  
- 0.5m(m-1)(2^{G1}-1)(9\cdot2^{G1}m-7m+2)\,,  
\end{equation}
where $N=3m(2^{G1}-1)+1$, $\langle l^2 \rangle$ is the mean-square bond length between units [1 for the cubic lattice used in the Monte Carlo (MC) simulations], and $G1$ is one more than the generation number.  	

For linear chains, $\langle S^2 \rangle$, there follows the well-known \cite{G} scaling law,				
\begin{equation} \label{e:s2}
  \langle S^2 \rangle = CN^{2\nu}.  
\end{equation}					
In this equation, the value of $C$ depends on the polymer model used, but the exponent, $2\nu$, is universal and is equal to 1 for ideal linear chains. Sheng et al. \cite{SJT} found that for a given generation, as the spacer length is increased, the size of an ideal dendrimer has the same exponent value. Earlier investigations by Burchard et  al. \cite{BKN} (see their figure 4) showed that for a fixed generation the radius of gyration scales linearly with the spacer length, but with a fixed spacer length it displays a logarithmic scaling. Equation~(\ref{e:s2}) was originally obtained in terms of the number of bonds, $N-1$, and for the finite systems studied in the Monte Carlo (MC) calculations this factor is used instead of $N$.
  
In order to measure the compactness of the dendrimer's structure, the $g$-ratio was calculated. This calculation involves the ratio of the radius of gyration of a dendrimer and the corresponding  linear polymer chain. Both polymers must contain the same number of units:
\begin{equation} \label{e:g}
          g = \frac {\langle S^2 \rangle} {\langle S^2 \rangle_l}  .
\end{equation}
There is an exact solution for the $g$-ratio determined by Wawrzyriska et al. \cite{WEPSZ} :
\begin{equation} \label{e:exactg}	
g = \frac {6(W_1 - W_2)}{(n/m)^3}.
\end{equation}
Here, $n$ is the number of bonds and
\begin{eqnarray} \label{e:W1}
W_1 &=& \frac {F_1}{(F_2-1)^3}\left[ F_2^{2G}(F_1F_2G-F_1G-F_1-F_2) \nonumber \right.\\ 
&+& \left. F_2^G(2F_1+F_1G-F_1F_2G-G+GF_2^2)-F_1+F_2\right] ,
\end{eqnarray}
whereas
\begin{equation} \label{e:W_2}	
W_2 = \frac {F_1^2(F_2^G-1)^2}{2(F_2-1)^2}+\frac {F_1(F_2^G-1)} {3(F_2-1)}.
\end{equation}
The tri-functional dendrimers have $ F_1=3$ and $F_2=2$.

The overall shape of the dendrimer can be found using the radius of gyration tensor. This tensor is composed of different eigenvalues. In three dimensions, they are $\lambda_3 \leqslant \lambda_2 \leqslant \lambda_1$. These eigenvalues are the principal moments of gyration, and there is one for each direction along the principal orthogonal axes \cite{SS}.  $\langle S^2 \rangle$ is equal to the average trace of the radius of gyration tensor, $\lambda_1 + \lambda_2 + \lambda_3$. The average asphericity $\langle A \rangle$ of a dendrimer in three dimensions is defined by Rudnick and Gaspari \cite{RG1,RG2} as
\begin{equation} \label{e:a2}
 \langle A \rangle   =   \left\langle \frac { \sum_{i>j}^3( \lambda_i -\lambda_j)^2 } { 2 (\sum_{i=1}^3\lambda_i)^2}\right\rangle,
\end{equation}
$\langle A \rangle$ is a number between 0 and 1, with 0 representing a perfectly spherical object, while an asphericity of 1 indicates that the units form a rod shape.  	
The prolateness,  $\langle P \rangle$, is also used to identify the overall shape. It is defined as
\begin{equation} \label{e:P}
   \langle P \rangle = \left\langle\frac {27(\lambda_1-\overline{\lambda})(\lambda_2-\overline{\lambda})(\lambda_3-\overline{\lambda})} {(\sum_{i=1}^3 \lambda_i)^3}\right\rangle,
\end{equation}
where $\overline{\lambda}$ is
\begin{equation} \label{e:av}
\overline{\lambda} = \frac {\lambda_1+\lambda_2+\lambda_3} {3}.
\end{equation}


It should be noted that $\langle S^2 \rangle$, the  $g$-ratio$, \langle A \rangle $, and  $\langle P \rangle $ are all rotationally
invariant universal quantities which can be expressed \cite{BJ} in terms of the trace, Tr, and determinant, Det, of the gyration 
tensor,\textbf{ Q}.
\begin{equation} \label{e:Q}
   Q_{ij} = \frac {1}{N} \sum_n^N (X_n^i-X_{CM}^i)  (X_n^j-X_{CM}^j)  \ \ \  i,j = 1...3.   
\end{equation}
Then,
\begin{equation} \label{e:Str}
\langle S^2 \rangle = \langle \Tr \ \textbf{Q} \rangle,
\end{equation}
and	
\begin{equation} \label{e:Atr}
\langle A \rangle =   \left\langle \dfrac {3   \Tr  \left(\textbf{Q}-\overline{\lambda} \textbf{I}\right)^2} {2  (\Tr \ \textbf{Q})^2} \right\rangle,
\end{equation}
where $\textbf{I}$ is the identity matrix.
Also,
\begin{equation} \label{e:Ptr}
\langle P \rangle =   \left\langle \dfrac {27 \Det  \left(\textbf{Q}-\overline{\lambda} \textbf{I}\right)} {(\Tr \  \textbf{Q})^3} \right\rangle.
\end{equation}


 Another important polymer property is the scattering function \cite{F2}, $S(k)$.
 Formally, this is defined as
\begin{equation}\label{e:sk}
  S(k) = \frac {1} {N^2}
\sum_{l}^N \sum_m^N  \re^{ \ri \mathbf{k} \ \bullet \ (\mathbf{R}_m - \mathbf{R}_l)}\,.
\end{equation}

Here, $N$ is the number of units in the polymer, $\mathbf{k}$ is the scattering vector and $\mathbf{R_l}$ and $\mathbf{R_m}$ are the respective positions of the $l$-th and $m$-th units. After averaging over the angles in three dimensions, the scattering function becomes
\begin{equation} \label{e:3ds(k)}
S(k) = \frac {1}{N^2} \left\langle\sum_{l}^N \sum_m^N \dfrac  {\sin x } {x}\right\rangle,
\end{equation}
where $x = k^2 \langle S^2 \rangle$. In the case of linear polymers, the theoretical values for the scattering function follow the Debye equation \cite{G},
\begin{equation} \label{e:debye}
S(k) = \frac {2(x -1+ \re^{-x})} {x^2}.
\end{equation}
Dendrimers are not linear and each structure will have its own equation for $S(k)$. The linear equation is included in one of the figures for comparison with the behavior of the dendrimers.

Dendrimers have received a lot of attention \cite{BHM} because they have applications in medical research, such as in pharmaceutical drug development. Their branched structure permits these molecules to be folded up into cages which can carry drugs into the body. In earlier work de Regt et al. \cite{RFBH} studied the properties of low generation tri-functional ideal structures containing nine or twenty-one branches. In this paper we extend their investigations to three higher generations, containing forty-five, ninety-three, and one hundred and eighty-nine branches, respectively.

\section{Methods}
\label{sec:headings}

Universal ratios for Gaussian (ideal) tree-branched polymers such as the $g$-ratio, $\langle A \rangle$, and $\langle P \rangle$  were computed with methods developed by Wei \cite{W1,W2} using the Kirchhoff matrix and its corresponding eigenvalues. These ratios are determined by extrapolation to infinite $N$. 


To calculate these shape parameters, the $N-1$ non-zero eigenvalues $\lambda_1, \ldots, \lambda_{N-1}$, of the $N\times N$ Kirchhoff matrix $\hat{K}$ for a random walk polymer with $N$ monomers is used whereby the characteristic polynomial of $\hat{K}$ can be expressed as

\begin{equation}
P_N(x)=\text{Det}[\Lambda_N(-x)]=\prod_{j=1}^{N-1}(\lambda_j-x)  .
\end{equation}
 From here the following functions; $D_N(x)$, $S_{1,N}(x)$ and $S_{2,N}(x)$ can be defined which are essential for the calculation of the shape parameters $\langle A \rangle$, $\langle P \rangle$ and the $g$-ratio. Using the notation $y=x/N$,

\begin{equation}
D_N(x) = \text{Det}[\Lambda_N^{-1}(0)]\text{Det}[\Lambda_N(y^2)] 
 = P_N^{-1}(0)P_{N}\left( -\frac{x^2}{N^2}\right) 
= \prod_{j=1}^{N-1}\frac{\lambda_j + y^2}{\lambda_j} \ ,
\end{equation}
and 
\begin{equation}
S_{1,N}(x) = \frac{1}{N^2}\text{\Tr}\left[\Lambda_N^{-1}(y^2)\right] = \frac{1}{N^2}\sum_{j=1}^{N-1}(\lambda_j +y^2)^{-1} \ .
\end{equation}
 Following the work of Wei, the functions for $S_{k,N}$ for $k = 2,3,..$ correspond to traces of higher powers of $\Lambda_N^{-1}$,

\begin{equation}
S_{2,N}(x) = \frac{1}{N^4}\sum_{j=1}^{N-1}\left(\lambda_j +y^2\right)^{-2} \ ,
\end{equation}
and
\begin{equation}
S_{3,N}(x) = \frac{1}{N^6}\sum_{j=1}^{N-1}\left(\lambda_j +y^2\right)^{-3} \ .
\end{equation}
The asphericity can then also be defined as
\begin{equation}
\langle A \rangle = \frac{15}{2}\int_0^{\infty}x^3 D_N^{-3/2}(x) S_{2,N}(x){\rd}x
= \frac{15}{2} \int_0^{\infty} \sum_{j=1}^{N-1}
\frac{y^3}{(\lambda_{j}+y^2)^2}\left[\prod_{k=1}^{N-1}\frac{\lambda_{k}}{\lambda_{k}+y^2}\right]^{3/2}{\rd}y
\end{equation}
and the prolateness as

\begin{equation}
\langle P \rangle = \frac{105}{8}\int_0^{\infty}x^5 D_N^{-3/2}(x) S_{3,N}(x){\rd}x
= \frac{105}{8} \int_0^{\infty} \sum_{j=1}^{N-1}
\frac{y^5}{(\lambda_{j}+y^2)^3}\left[\prod_{k=1}^{N-1}\frac{\lambda_{k}}{\lambda_{k}+y^2}\right]^{3/2}{\rd}y.
\end{equation}
The $g$-ratio is given as

\begin{equation}
g = S_{1,N}^{\text{dendrimer}}(0)/S_{1,N}^{\text{linear}}(0) 
 = \sum_{j=1}^{N-1}\lambda_j^{-1}/\sum_{k=1}^{N-1}\tilde{\lambda}_k^{-1} , 
\end{equation}
where $\lambda_j$ and $\tilde{\lambda}_k$ are the non-zero eigenvalues of the Kirchhoff matrix for a dendrimer and linear polymer with $N$ monomers.


The exact 9-branch and 21-branch dendrimer $S(k)$'s have previously been determined \cite{RFBH}:
\begin{equation} \label{e:9branch}
S(k)_9 = \frac {2}{x^2}\left(3 + x -3\re^{-x/9}-12\re^{-x/3}+12\re^{-4x/9}\right)
\end{equation}
and
\begin{equation} \label{e:21branch}
S(k)_{21} = \frac{2} {x^2} \left(9 + x - 3\re^{-x/21} + 6\re^{-2x/21} - 12\re^{-x/7} - 48\re^{-5x/21} + 48\re^{-2x/7}\right) .
\end{equation}

We have applied the general algorithm detailed in the Appendix in combination with the Benhamou et al. \cite{BGGB} formalism  to obtain results for 45, 93 and 189-branched systems. Their method expresses $S(k)$ as

\begin{equation} \label{e:BGGB}
S(k) = \dfrac{1}{f} \left (x -1 + \re^{-x} + \dfrac {2(\re^{2x} - 2\re^x +1)}  {fx^2} \sum_{k \subset G,|k| \geqslant 2} \re^{-|k| x} \right).
\end{equation}
Here, $k$ is the length of a connected path between any two junctions in the graph $G$ and
 $ x = k^2 \langle S^2 \rangle $ calculated for the linear chain. The algorithm developed in the Appendix first determines the number of walks of
length $k$ connecting two junctions and then prunes this set to obtain the number of paths. A walk may visit a junction more than
once whereas a path cannot. 

\begin{table}[!b]
\caption{Number of paths $p^f_k$ of length $k$ in an $f$-branch dendrimer.}
\label{table1}
\begin{center}
\renewcommand{\arraystretch}{0}
\begin{tabular}{c|c|c|c|c|c|c|c|c|c|c|c}
\hline\hline 
\strut$k$ & 2 & 3 & 4 & 5 & 6 & 7 & 8 & 9 & 10 & 11 & 12\\
\hline \strut$p^{45}_k$ & 66 & 84 & 120& 144& 192& 192& 192 & 0 & 0 & 0 & 0 \\
\hline\strut $p^{93}_k$ & 138 & 180 & 264 & 336 & 480 & 576 & 768 & 768 & 768 & 0 & 0 \\
\hline\strut $p^{189}_k$ & 282 & 372 & 552 & 720 & 1056 & 1344 & 1920 & 2304 & 3072 & 3072 & 3072 \\
\hline\hline
\end{tabular}
\renewcommand{\arraystretch}{1}
\end{center}
\end{table}

Using the data in table \ref{table1} (see the Appendix for how these numbers were obtained) we find that
\begin{eqnarray}  \label{e:45branch}
S(k)_{45} &=& \dfrac{2}{x^2}\left(21+x - 3\re^{-x/45} +18\re^{-2x/45} -12\re^{-x/15} + 24\re^{-4x/45}  \nonumber\right.\\
&-&\left. 48\re^{-x/9}-192\re^{-7x/45} +192\re^{-8x/45}\right).
\end{eqnarray}
This has a first-order Taylor expansion 
\begin{equation}
S(k)_{45} = 1 - \frac{3089x}{30375} + ...\,.
\end{equation}
The $g$-ratio can then be determined by dividing the linear term in $x$ from $S(k)_{45}$ by the corresponding term in  the Taylor expansion for a linear chain:
\begin{equation}
S(k)_{\ell} = 1 - \frac{x}{3} + ...\,.
\end{equation} 
This gives a $g$-ratio value of $3089/10125 \approx 0.3051$.

The $S(k)$ expressions for 93 and 189 branch dendrimers are
\begin{eqnarray}  \label{e:93branch}
S(k)_{93} &=& \dfrac{2}{x^2}\left(45 + x -3\re^{-x/93} + 42\re^{-2x/93} -12\re^{-x/31} +72\re^{-4x/93}\nonumber\right.\\
&-&\left. 48\re^{-5x/93} +96\re^{-2x/31} -192\re^{-7x/93} -768\re^{-3x/31}
+768\re^{-10x/93}\right)
\end{eqnarray}
and
\begin{eqnarray}  \label{e:189branch}
S(k)_{189} &=& \dfrac{2}{x^2}\left(93 + x -3\re^{-x/189} +90\re^{-2x/189} -12\re^{-x/63} +168\re^{-4x/189}\right. \nonumber\\
&-&48\re^{-5x/189} +288\re^{-2x/63} -192\re^{-x/27} +384\re^{-8x/189}
-768\re^{-x/21} \nonumber\\
&-&\left.3072\re^{-11x/189}+3072\re^{-4x/63}\right).
\end{eqnarray}

After Taylor expanding and dividing by the linear chain result, we find that the $g$-ratios for 93 and 189 branched structures are approximately 0.2009 and 0.1271,  respectively.

The MC simulations employed the following algorithm to generate random configurations. The dendrimers are grown on a three-dimensional simple cubic lattice. The central unit is placed at the origin of the coordinate system. Then, a random number is generated to select one of the six possible directions for placing the second unit a distance one apart from the central unit.  Each successive placement of units is performed by this procedure. Since ideal dendrimers are the focus of this research, the units are allowed to overlap.  Each completed dendrimer is an independent configuration and properties were computed by averaging over each of these random samples. In all cases, 100~000 random samples were employed.

\section{Results}
\label{sec:headings}

Tables \ref{table2}, \ref{table3}, and \ref{table4} show the property values found for the different branch dendrimers as a function of the number of units $N$; $\langle \lambda \rangle$ values for each of the eigenvalues are included, as well as, $\langle A \rangle$ and $\langle P \rangle$. The radius of gyration is presented for both the dendrimer and the corresponding linear polymer chain. The error deviation is shown as a number in parenthesis (one standard deviation from the mean).

The tables contain some interesting results. For any dendrimer generation, $\langle A \rangle$ and $\langle P \rangle$ level off as the number of total units in the dendrimer increases. Another observation is that $\langle S^2 \rangle$ is always higher for the linear chains than it is for the dendrimers. This is to be expected because the dendrimer is less spread out than a linear chain and would therefore have an average shorter distance from its center of mass to an outer unit.

 \begin{table}[!t]
\caption{Effect of the number of units, $N$, for 45-branch dendrimers.}
\label{table2}
\vspace{2ex}
\begin{center}
\begin{tabular}{ccccccc}
\hline\hline\strut Property/$N$ & 901 & 1531 & 1756 & 2206  \\
\hline
$\langle \lambda_1 \rangle$ &26.79(3)&45.45(5)&52.11(6)&65.54(8)     \\
$\langle \lambda_2 \rangle$ &12.78(1)&	21.60(2)&	24.81(3)	&31.09(3) \\
$\langle \lambda_3 \rangle$ &6.57(1)&	11.09(1)&	12.70(1)&	15.93(2)   \\
$\langle A \rangle$ &	0.160(1)&	0.161(1)&	0.161(1)&	0.162(1)   \\ 
$\langle S^2 \rangle$ &46.14(4)	&78.13(7)&	89.62(8)&	112.56(10) \\
$\langle P \rangle$ &0.095(1)	&0.096(1)&	0.096(1)	&0.097(1)   \\ 
$\langle S^2 \rangle_{\ell}$ &149.51(24)	&254.05(42)	&291.51(47)	&366.89(60)	 \\ 
\hline\hline 
\end{tabular}
\end{center}
\end{table}

\begin{table}[!t]
\caption{Effect of the number of units, $N$, for 93-branch dendrimers.}
\label{table3}
\vspace{2ex}
\begin{center}
\begin{tabular}{ccccccc}
\hline\hline\strut Property/$N$ & 931 & 1861 & 2791 & 3721  \\
\hline
$\langle \lambda_1 \rangle$ &17.23(2)&	34.28(4)&	51.36(5)	&68.39(7)    \\
$\langle \lambda_2 \rangle$ &9.15(1)&	18.12(2)&	27.11(2)	&36.12(3) \\
$\langle \lambda_3 \rangle$ &5.23(1)&	10.32(1)&	15.44(1)&	20.54(2)   \\
$\langle A \rangle$  &	0.121(1)&	0.123(1)&	0.123(1)&	0.123(1)   \\ 
$\langle S^2 \rangle$ &31.61(2)	&62.73(5)&	93.91(7)&	125.05(10) \\
$\langle P \rangle$ &	0.059(1)	&0.060 (1)&	0.060(1)&	0.060(1)   \\ 
$\langle S^2 \rangle_{\ell}$ &154.54(25)&	309.18(50)&	465.57(76)&	619.62(101)	 \\ 
\hline\hline 
\end{tabular}
\end{center}
\end{table}

\begin{table}[!b]
\caption{Effect of the number of units, $N$, for 189-branch dendrimers.}
\label{table4}
\vspace{2ex}
\begin{center}
\begin{tabular}{ccccccc}
\hline\hline\strut Property$/N$ & 946 & 2080 & 3592 & 4915  \\
\hline
$\langle \lambda_1 \rangle$ 	&10.53(1)&	22.94(2)&	39.55(4)	&54.10(5)     \\
$\langle \lambda_2 \rangle$ &6.12(5)&	13.26(1)&	22.76(2)&	31.12(2) \\
$\langle \lambda_3 \rangle$ &3.83(1)&	8.26(1)&	14.14(1)&	19.32(1)   \\
$\langle A \rangle$ &	0.091(1)&	0.092(1)&	0.093(1)&	0.094(1)	  \\ 
$\langle S^2 \rangle$ 	&20.48(1)&	44.47(3)	&76.45(5)	&104.53(7) \\
$\langle P \rangle$ &	0.036(1)&	0.037(1)&	0.038(1)&	0.038(1) \\ 
$\langle S^2 \rangle_{\ell}$ &	157.12(26)	&345.21(56)&	599.47(98)&	819.05(134) \\ 
\hline\hline 
\end{tabular}
\end{center}
\end{table}

\begin{figure}[!t]
\centerline{\includegraphics[width=0.55\textwidth]{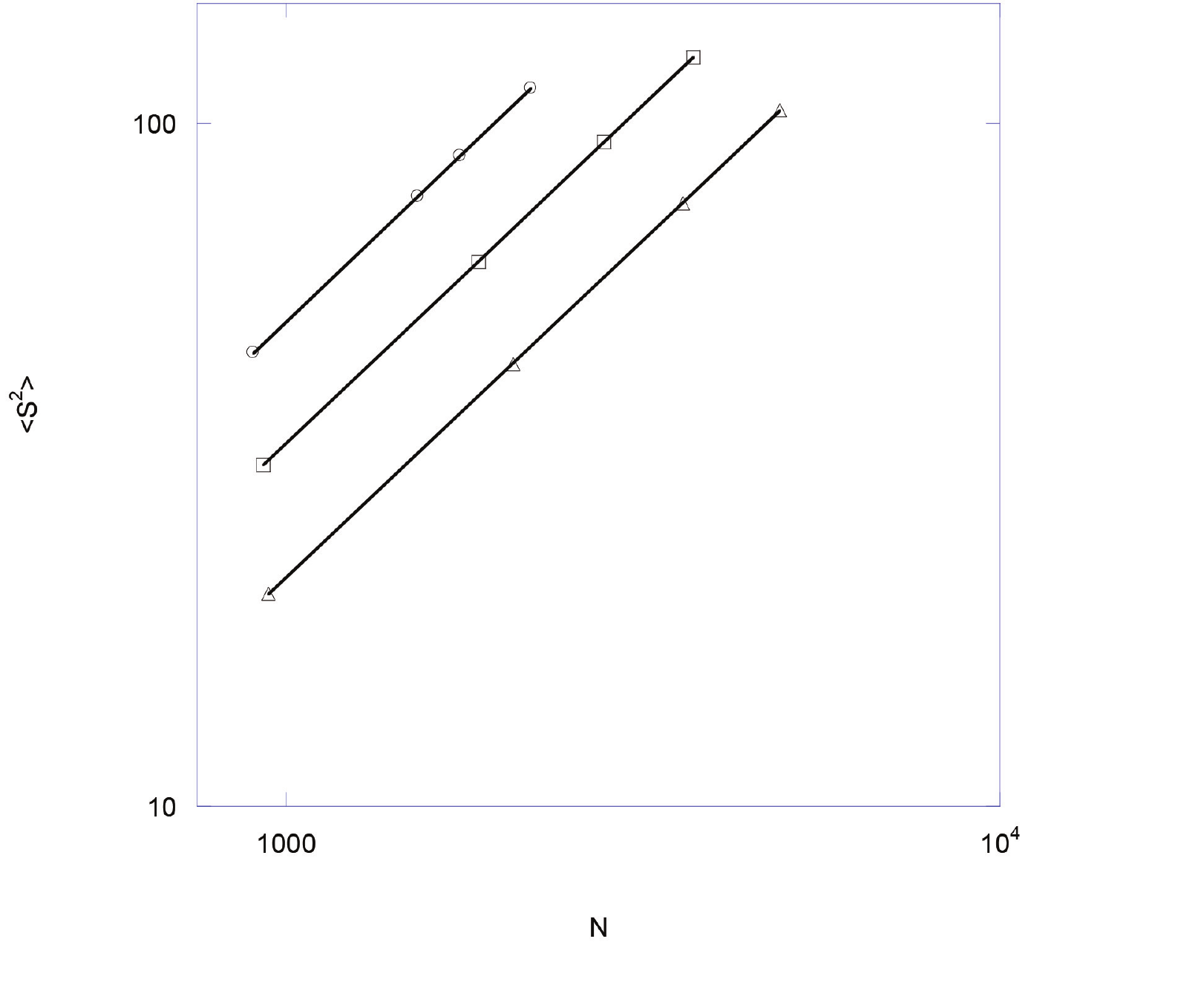}}
\caption{Log-log plot of the radius of gyration vs. $N$ for each dendrimer generation: circles 45-branch, squares 93-branch, triangles 189-branch, lines best linear fit.} \label{loglog}
\end{figure}

Table \ref{table5} demonstrates that the radius of gyration data reported in tables \ref{table2}, \ref{table3}, and \ref{table4} is in excellent agreement with the Sheng et al. \cite{SJT} exact equation. The scaling exponent for the radii of gyration, $2\nu$, was obtained by fitting the $\langle S^2 \rangle$ data to a power function. Figure \ref{loglog} presents a log-log plot of the data for each distinct generation. The exponents found are:  0.995(1), 0.992(1), and 0.988(1) for 45, 93 and 189 branches, respectively. These are close to the expected theoretical exponent of 1.0.

\begin{table}[!t]
\caption{Comparison of $\langle S^2 \rangle$ MC to Sheng et al.\cite{SJT} exact predictions.}
\label{table5}
\begin{center}
\begin{tabular}{ccccccccc}\cline{1-3}\cline{1-3} \cline{1-3}
\hline
\multicolumn{3}{c}{45 Branches}&
\multicolumn{3}{c}{93 Branches}&
\multicolumn{3}{c}{189 Branches}
\\
\hline
\cline{1-3} 
\cline{1-3} \cline{1-3}
$N$ & 
MC &  
Exact  & 
 $N$ & 
  MC &  
   Exact &	 
   $N$ & 
    MC & Exact \\
\hline 
901
&
46.14(4)&
46.161
&
931&
31.61(2)
&
31.578&
946
&
20.48(1)&
20.469 \\             
1531
&78.13(7)&78.195
&1861&62.73(5)
&62.723&2080
&44.47(3)&44.484 \\            
1756
&89.62(8)&89.636
&2791&93.91(7)
&93.868&3592
&76.45(5)&76.503\\           
2206
&112.56(10)&112.517
&3721&125.05(10)
&125.014&4915
&104.53(7)&104.520\\
\hline              
\cline{1-3} 
\cline{1-3} \cline{1-3} 
\end{tabular}
\end{center}
\end{table}

The data in the tables are for finite $N$ whereas most of the theories are for infinite $N$. Thus, the data for $\langle A \rangle$ and $\langle P \rangle$ have been extrapolated using a linear function in $1/N$. The $g$-ratios were extrapolated after first determining their error by relating the error in a ratio to the error in the numerator and the error in the denominator.  As the number of units increases, $1/N$ gets smaller, so that when $1/N \rightarrow 0$, $N$ becomes infinite. Hence, extrapolation should remove all finite $N$ effects. 


Table \ref{table6} compares the extrapolation MC $g$-ratios to Eq.~(1.6), the Wei and BGGB methods.  All the results are in excellent agreement.
The MC extrapolations for  $\langle A \rangle$ and  $\langle P \rangle$ are compared to the Wei method in table \ref{table7}. Again excellent agreement is found.


\begin{table}[!t]
\caption{Comparsion of extrapolated MC $g$-ratios to theoretical predictions.}
\label{table6}
\vspace{2ex}
\begin{center}
\begin{tabular}{ccccc}
\hline\hline\strut
 Branches & MC & Eq. 1.6 & Wei & BGGB  \\
\hline
45 &0.306(2)& 0.305086&0.305088&0.3051  \\  
93 &0.201(1) & 0.200937&0.200938&0.2009  \\  
189 &0.128(1)& 0.127060&0.127062&0.1271 \\                             
\hline\hline 
\end{tabular}
\end{center}
\end{table}
																			
\begin{table}[!t]
\caption{Comparison of  extrapolated MC $\langle A \rangle$ and $\langle P \rangle$ to Wei method results.}
\label{table7}
\vspace{2ex}
\begin{center}
\renewcommand{\arraystretch}{0}
\begin{tabular}{ccccc}
\hline\hline\strut Branches &$\langle A \rangle$ MC  & Wei&$\langle P \rangle$  MC  & Wei\\
\hline   \strut          
45 &0.163(2) &0.1623100&0.098(1)& $0.0967390^a$\\
93 &0.124(1)&0.1237990&0.061(1)& $0.0609874^a$\\
189&0.094(1)&0.0944898&0.039(1)&$0.0389064^a$  \\                                       
\hline\hline
\end{tabular}
\renewcommand{\arraystretch}{1}
\end{center}
{\raggedright \centering $a$ Note that the Wei method defines $\langle P \rangle$ with an additional factor of $1/2$. \par}
\end{table}

The values of the asphericity reflect how spherical and symmetric a dendrimer is; an $\langle A \rangle$ value of 0.0 represents a perfect sphere. The data show that, as the number of branches increases, the values for $\langle A \rangle$ and $\langle P \rangle$ decrease. De Regt et al. \cite{RFBH} found that $\langle A \rangle$ changed from 0.266(1) to 0.211(1) when going from the first generation 9-branch dendrimer to the second generation 21-branch dendrimer; similarly $\langle P \rangle$ changed from 0.233(3) to 0.153(2). These values indicate that higher generation dendrimers have a more symmetric and spherical shape.  

In table \ref{table8} the $\langle A \rangle$ and $g$-ratio results for ideal dendrimers are compared to  Wawrzyriska et al. \cite{WSZ} values for a MC self-avoiding walk (SAW) model. One notes that the excluded volume effects in the SAW become more and more important as the generation number increases. Indeed, the excluded volume model is forced to be more symmetrical because the units cannot overlap. 


It is interesting to compare the shape properties of dendrimers to multi-branched comb polymers. Casassa and Berry \cite{CB} have derived a general equation for the $g$-ratio of uniform three-functional ideal comb polymers:
\begin{equation}
g_\text{comb} = r - \dfrac{r^2(1-r)} {f+1} + \dfrac{2r(1-r)^2} {f}+\dfrac{(3f-2)(1-r)^3} {f^2},
\end{equation} 
where $r$ is the ratio of the number of branches in the backbone to the total number of branches.
This equation yields 0.638, 0.566, 0.532, 0.516 and 0.508 when $f$ is 9, 21, 45, 93 and 189, respectively.
Comparing these values to the dendrimer results in table \ref{table8} one sees a very different behavior. The $g$-ratios of the
comb polymers appear to  level off to 0.5 as the number of branches increase whereas the dendrimer $g$-ratios
get smaller and smaller.  Van Ferber et al \cite{FBFRZ} have computed $\langle A \rangle$ for an ideal 9 branch  comb polymer. They found
that  $\langle A \rangle$ = 0.295(2) which is larger than the 9 branch dendrimer value of 0.266(1). The $g$-ratio and $\langle A \rangle$ value indicate that dendrimers are more symmetrical than the corresponding comb polymers.

\begin{table}[!t]
\caption{Comparison of ideal and SAW dendrimers.}
\label{table8}
\vspace{2ex}
\begin{center}
\renewcommand{\arraystretch}{0}
\begin{tabular}{ccccc}
\hline\hline\strut Branches & Ideal $\langle A \rangle$ & $g$-ratio & SAW $\langle A \rangle$ & $g$-ratio  \\
\hline\strut
9   &0.266(1)$^a$ &0.606(1)$^a $&	0.2682	 &0.6068     \\
21  &0.211(1)$^a$ &0.443(1)$^a $&0.1968  &0.4485	 \\
45  &0.163(2)    &0.306(2)     & 0.1390  &0.3180	   \\
93  &0.124(1)    &0.201(1)     & 0.0972  &0.2199	  \\ 
189 &0.094(1)    &0.128(1)     & 0.0685  &0.1509 \\
\hline\hline 
\end{tabular}
\renewcommand{\arraystretch}{1}
\end{center}
\vspace{-0.2cm}
{\raggedright\centering $a$ see reference \cite{RFBH}. \par}
\end{table}
\begin{figure}[!b]
\centerline{\includegraphics[width=0.60\textwidth]{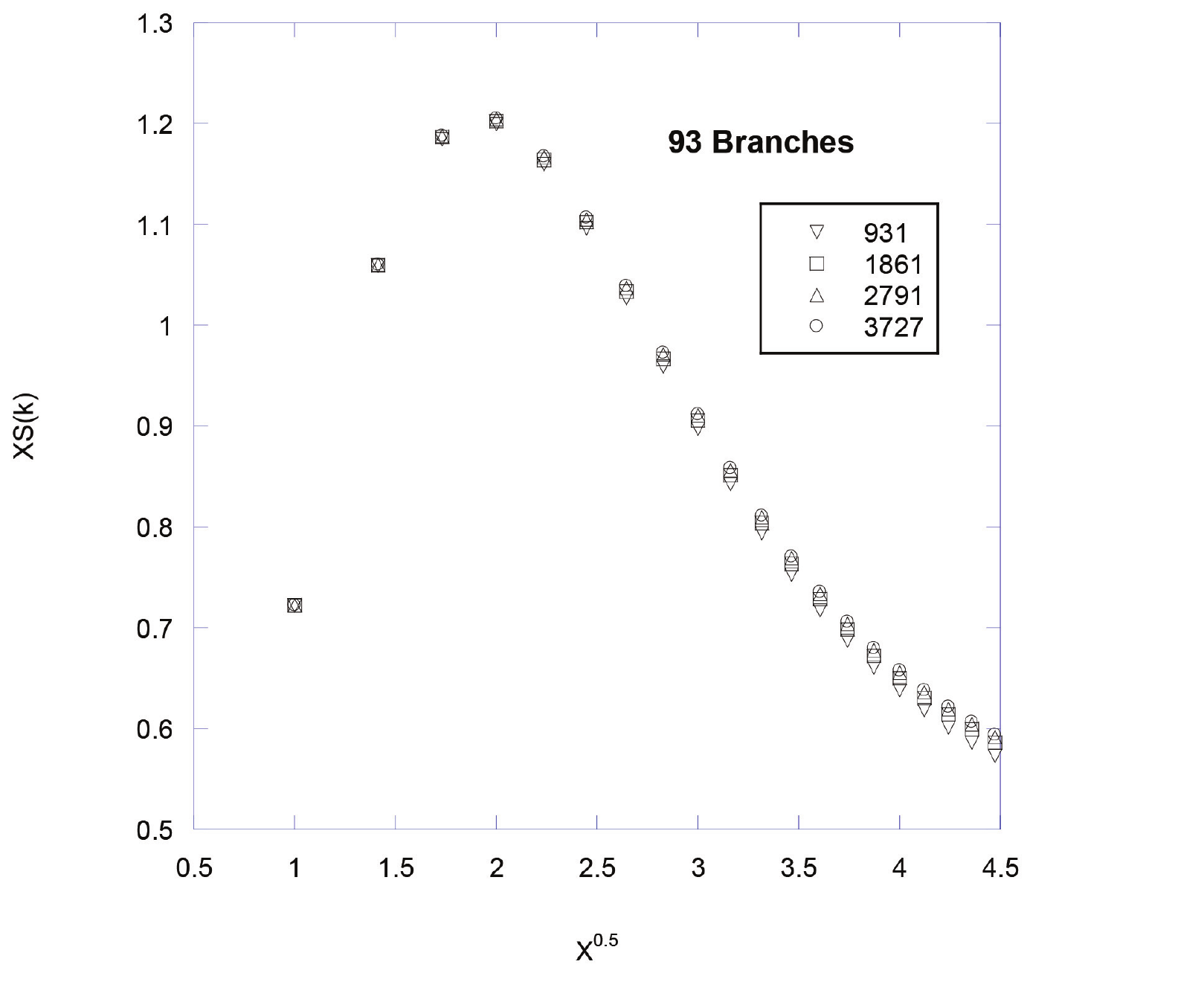}}
\caption{Number dependence for the 93-branch dendrimer scattering function.} \label{diagnumb}
\end{figure}

A Kratky plot \cite{B} of $S(k)$ is often used to investigate branched polymers. Figure \ref{diagnumb} shows the variation of a Kratky plot for the MC 93-branch dendrimer $S(k)$ calculations as the total number of units in all the branches is increased from 931 to 3727. Note that we have used  $ x = k^2 \langle S^2 \rangle $ throughout this paper instead of the usual $x^2= k^2 \langle S^2 \rangle $ to conform to the  notation employed by  Benhamou et al. \cite{BGGB}.  It is clear that number dependent effects are quite small if one uses a sufficiently large number of units in the simulations.

\begin{figure}[!t]
\centerline{\includegraphics [width=0.70\textwidth]{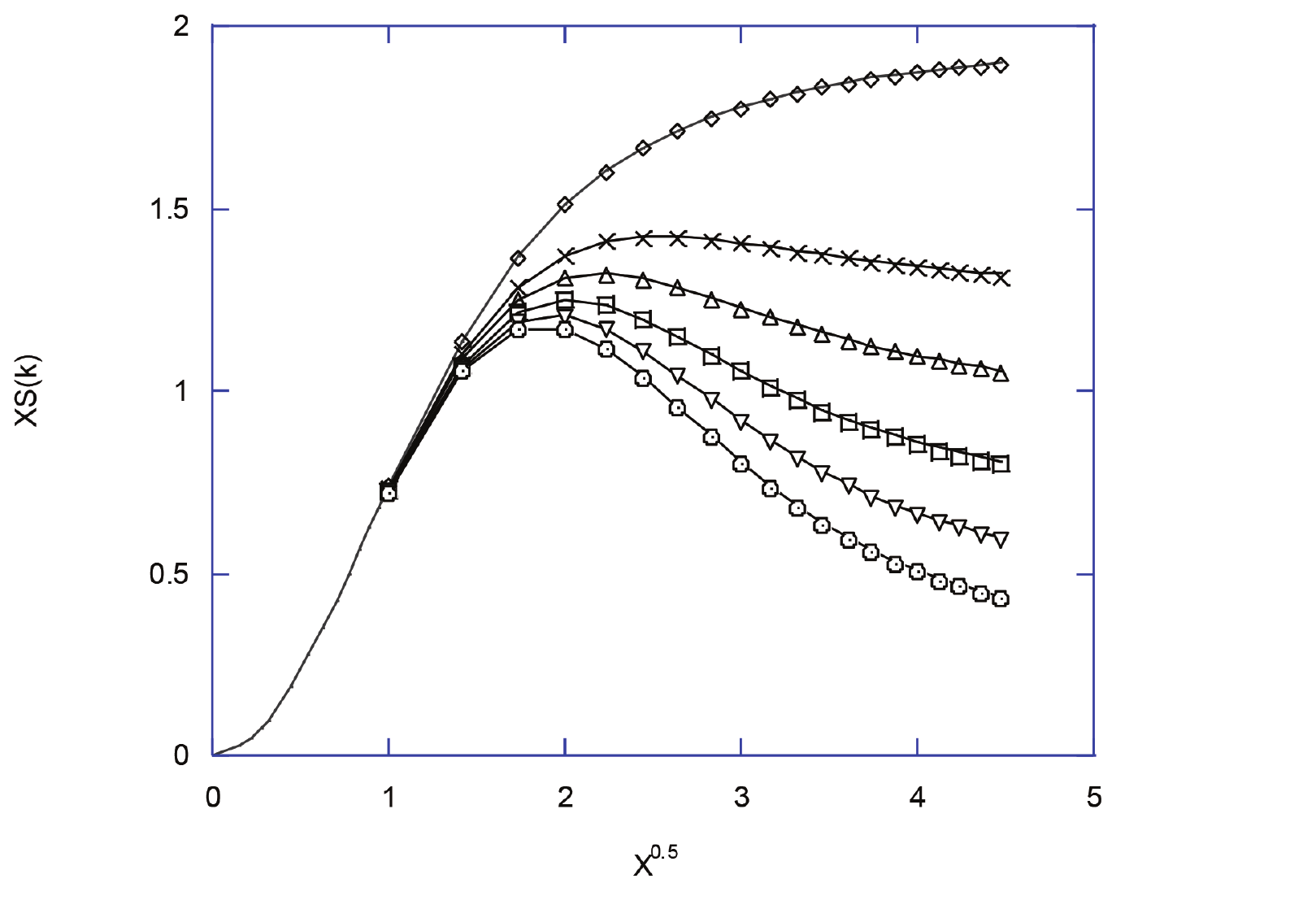}}
\caption{Variation of the scattering function with dendrimer generation and number of units: diamond linear 2080, X 9-branch 1342, up triangle  21-branch 3130, square 45-branch 2206, down triangle 93-branch 3727, circle 189-branch 4915, lines exact equations.} \label{skplot}
\end{figure}

Figure \ref{skplot} presents the results for the largest $N$, $S(k)$  MC calculations for linear chains and 9, 21, 45, 93 and 189-branch dendrimers compared to the exact equations. In order to compare the structures with different branching arrangements, the data have been scaled with the appropriate $g$-ratio values as described in Toporowski and Roovers \cite{TR}.  As reported by Burchard \cite{B}, ``Linear randomly coiled chains result in an angular asymptote which has a value of 2 for monodisperse coils\ldots{} The striking feature with branched chains is the appearance of a maximum for stars and other regularly branched chains. This maximum becomes more and more pronounced with an increasing branching density.''  This plot confirms the observation that higher generation dendrimers are more symmetrical. Even higher dendrimer generations are needed to see sphere-like behavior \cite{GB}.

\section{Conclusions}
 


Three different ideal dendrimer generations have been examined using a variety of methods. Multiple properties were computed, and excellent agreement was found between the Monte Carlo simulations and the theoretical predictions. All the properties studied demonstrate that ideal dendrimers become more symmetrical as the generation number and the number of branches increase.

As we have shown,  excluded volume effects become more and more important at higher generations. Coarse grained simulation models reviewed by Klos and Sommer \cite{KS} have established that the radius of gyration of isolated excluded volume dendrimers has different scaling behaviors in good, theta and poor solvents. Even more realistic models of isolated charged dendrimers have demonstrated the important influence of parameters such as pH on the structural properties. The ideal systems examined here form a first approximation to these more complicated models and permit exact calculations for many dendrimer properties.


\section*{Acknowledgements}

We wish to thank the Manhattan College Computer Center and the Kakos Center for Scientific Computing for large grants of computer time. We would also like to thank Professors Mark DeBonis and  Paula Whitlock for helpful discussions. Brandon Thrope was supported by a Manhattan College Summer Research grant.


\section*{Appendix}

The structure of a dendrimer can be considered as an abstract graph, $G$, in which the polymer junctions are represented by nodes (vertices, $V$) and the branches as edges, $E$.  A \textit{walk} in such a graph $G = (V,E)$ is then a sequence of nodes $v_1, v_2, \ldots v_n$ so that $E(v_i,v_{i+1})$ for each $i$ with $1 \leqslant i \leqslant n-1$.  A \textit{path} in $G$ is a walk in which the nodes are pairwise distinct (so, a walk may visit a node more than once, while a path may not).

In order to be able to compute the theoretical $S(k)$ and thus the $g$-ratio for a given polymer $T$, we need to be able to count the number of paths of each length in $T$.  Since each of the polymers considered here are dendrimer polymers (having a tree structure), we know that between any two nodes, there is a unique path (trees have no cycles, i.e., they have no paths which begin and end at the same node).  Therefore, if $T$ is a dendrimer polymer with $N$ nodes, then $T$ has exactly $N \choose 2$ total paths.  However, we are interested in finding the number of paths of any given length.

 We define the \textit{adjacency matrix} $A$ of the graph $G$ with node set $V = \{1, \ldots, N\}$ to be the $N \times N$ matrix $(a_{ij})$ so that $a_{ij} = 1$ if $\{i,j\} \in E$, and $a_{ij} = 0$ if $\{i,j\} \not\in E$, where $a_{ij}$ denotes the entry in the $i$-th row and $j$-th column of the matrix.  Let $M = (m_{ij})$ be a matrix.  We denote $M(i,j) = m_{ij}$ for each $i,j$, and $M^n$ to be the $n$-th power of the matrix $M$ (i.e., $M$ multiplied by itself $n$-many times).  From \cite{RA}, we see that the number of walks in $G$ from node $i$ to node $j$ of length $k$ is given by $A^k(i,j)$ (i.e., the entry in the matrix $A^k$ contained in the $i$-th row and $j$-th column).

Therefore, one way to count the number of paths of length $k \in \mathbb{N}$ in a given dendrimer polymer $T$ (of the type under consideration) with adjacency matrix $A$, would be to do the following:
	\begin{enumerate}
		\item Case $k=2$:  
			\begin{enumerate}
				\item Compute $A^2$ and initialize $A_2 = A^2$.
				\item Change all main diagonal entries in $A_2$ to 0.  
				\item Sum all entries in $A_2$.
				\item Divide the result by 2.
			\end{enumerate}
		\item Case $k>2$ with $k \in \mathbb{N}$: 
			\begin{enumerate}
				\item Compute $A^k$ and initialize $A_k = A^k$.  
				\item Change all entries of $A_k$ that are greater than $1$ to $0$.  
				\item Sum all entries in $A_k$.
				\item Divide the result by 2.
			\end{enumerate}
	\end{enumerate}
	
\textit{Verification:} 

\underline{Case $k=2$.}  Every main diagonal entry in $A^2$ will be greater than $0$, but these entries do not correspond to paths.  They correspond to walks from a node back to the same node, and therefore they are not valid paths; each of these entries therefore counts the degree of the corresponding node.  The degrees of each leaf node is $1$ (the \textit{degree} of a node is the number of nodes adjacent to it, and a \textit{leaf node} in a tree is a node that is only adjacent to one other node), and, therefore, there will be some main diagonal entries that are exactly $1$.  Since, in this case, we are counting length-two paths, no other entries in $A^2$ will be greater than 1.  It is easy to see that the non-main-diagonal entries of 1 in $A^2$ correspond to valid paths.

After changing the main diagonal entries to $0$, and summing the remaining entries, we will have double the number of valid paths, because if there is a path from $i$ to $j$, then there is also a path from $j$ to $i$ (they are the same path), so that we will have $A_2(i,j) = 1 = A_2(j,i)$.

\underline{Case $k>2$, $k \in \mathbb{N}$.}
We do not need to treat main diagonal entries separately in this case.  If $k$ is odd, then the main diagonal entries will all be $0$ (because there are no odd-length walks from a node back to itself), and if $k$ is even with $k>2$, then the main diagonal entries corresponding to leaf nodes will be greater than $1$, since the nodes adjacent to the leaf nodes have degree larger than $1$ (so, there will always be more than $1$ walk from every node back to itself).

First, we claim that if $A^k(i,j)=0$, then there is no path between $i$ and $j$.  Suppose $A^k(i,j)=0$.  Then, there is no walk between $i$ and $j$, and therefore there can be no path between $i$ and $j$, proving the claim.

Next, we claim that if $A^k(i,j) > 1$, then there is no path of length $k$ between $i$ and $j$.  Dendrimer polymers have a tree structure, in which there is not more than one distinct path between two given nodes.  So, suppose that $A^k(i,j) > 1$ and there is exactly one path between $i$ and $j$.  Let $p$ denote the path and let $w$ denote a walk between $i$ and $j$ that is different from $p$.  It cannot be that both $p$ and $w$ are paths, because then we could form a cycle, which is impossible because of the tree structure of the polymer.  Thus, $w$ must repeat some node.  Then, from this fact, we can argue that we could form some path $p'$ from $i$ to $j$ which has the length less than $k$.  Clearly, $p$ cannot be the same as $p'$, and therefore we could again form a cycle, which is impossible.  This proves the claim.

Now, we claim that if $A^k(i,j) = 1$, then there is exactly one valid path between $i$ and $j$ of length $k$.  Suppose that $A^k(i,j) = 1$, but there is not exactly one path between $i$ and $j$ of length $k$.  (There cannot be more than one such path because there is only one walk.)  Because $A^k(i,j) = 1$, there must be a unique walk $w$ (of length $k$), consisting, say, of nodes $\{v_1, \ldots, v_k\}$ with $v_1=i$ and $v_k=j$, which passes through some $v_q$, with $q\in \{1,\ldots,k\}$, more than once.  But then we can find a walk $w' \ne w$ from $i$ to $j$.  It cannot be that there is a cycle starting at $v_q$ because the polymer has a tree structure.  Therefore, it must be the case that $w$ passes from $v_q$, then through a line of nodes, then traverses back through the same line of nodes to reach $v_q$ once again.  Without loss, assume that $w$ has the form $v_1, \ldots, v_q, v_{q'}, v_q, \ldots, v_k$.  Because of the structure of the dendrimer polymers under consideration (and the fact that each non-leaf node has degree $3$), there must be some $v_{\ell}$ in $w$ which is adjacent to some node $v'$ which is not in $w$.   Define $w'$ to consist of nodes $v_1, \ldots, v_{\ell}, v', v_{\ell}, \ldots v_k$ (where $v_{q'}$ is excluded, and $v_{\ell}$ could be either $v_1$ or $v_k$ as well).  Thus, there is more than one distinct walk between $i$ and $j$, contradicting the assumption that $A^k(i,j) = 1$.  This proves the claim.

Finally, after changing the entries in $A^k$ which are larger than $1$ to $0$, and summing the remaining entries, we will have double the number of valid paths.

\textit{This ends the verification.}
	
The above algorithm was implemented in MATLAB for the dendrimer polymers under consideration.  If $p^f_k$ denotes the number of paths of length $k$ in the $f$-branch dendrimer polymer, we obtain the data presented in table \ref{table1} to use in equation~(\ref{e:BGGB}).

\ukrainianpart

\title[Дослідження ідеальних дендримерів вищого порядку методом Монте-Карло]%
{Дослідження ідеальних дендримерів вищого порядку методом Монте-Карло%
}
\author[М. Юра, M. Бішоп, Б. Троуп, Р. де Регт]{М.  Юра\refaddr{label1},
	М. Бішоп\refaddr{label1}, Б. Трофе\refaddr{label1},
	Р. де Регт\refaddr{label2, label3}}
\addresses{
	\addr{label1} Математичний факультет, Мангаттан Колледж, Мангаттан Колледж Парквей, Рівердейл, Нью-Йорк 10471, США
	\addr{label2} Центр прикладної математики, університет Ковентрі, Ковентрі CV1 5FB, Великобританія
	\addr{label3} Докторантський коледж статистичної фізики складних систем, Лейпціг-Лоран-Львів-Ковентрі (L4), D-04009 Лейпціг, Німеччина}

\makeukrtitle

\begin{abstract}
	Досліджено властивості ідеальних три-функціональних дендримерів з сорока п'ятьма, дев'яносто трьома та сто вісімдесят дев'ятьма гілками. Для розрахунку середньоквадратичного радіуса гірації, $g$-співвідношення, асферичності, параметрів форми та форм-фактора використовуються три різні методи. До них належать: техніка власних значень матриці Кірхгофа, 
	формалізм в рамках теорії графів Бенаму та ін. (2004), а також моделювання Монте-Карло з використанням алгоритму зростання. Запропоновано нову методику підрахунку траекторій при представленні дендримерів у вигляді графів. Усі методи відмінно узгоджуються між собою та з наявними теоретичними передбаченнями. Дендримери стають більш симетричними при збільшенні порядку гілок та їх кількості.
	\keywords м'яка речовина, дендример, аналітичний підхід, моделювання Монте-Карло
\end{abstract}


\begin{thebibliography} {99}
\bibitem{F1} Flory P. J., Principles of Polymer Chemistry, Cornell University Press, Ithaca, 1953.
\bibitem{SJT}Sheng Y. J., Jiang S., Tsao H. K., Macromolecules, 2002,  \textbf{35},  7865, \doi{10.1021/ma025561k}.
\bibitem{G}  De Gennes P. G., Scaling Concepts in Polymer Physics, Cornell University Press, Ithaca, 1979.
\bibitem{BKN}  Burchard W., Kajiwara K., Nerger D., J. Polym. Sci. Polym. Phys. Ed., 1982, \textbf{20},  157,\\ \doi{10.1002/pol.1982.180200201}.
\bibitem{WEPSZ} Wawrzyriska E., Eisenhaber S., Parzuchowski P.,  Sikorski A.,  Zifferer G., Macromol. Theory Simul., \\ 2014, \textbf{23},  288, \doi{10.1002/mats.201300159}.
\bibitem{SS}  Solc K., Stockmeyer W. H., J. Chem. Phys., 1971, \textbf{54}, 2756, \doi{10.1063/1.1675241}.
\bibitem{RG1}  Rudnick J., Gaspari G., Science, 1987, \textbf{237}, 384, \doi{10.1126/science.237.4813.384}.
\bibitem{RG2}  Rudnick J.,  Gaspari G., J. Phys. A, 1986, \textbf{19}, L191, \doi{10.1088/0305-4470/19/4/004}.
\bibitem{BJ} Blavatska V.,  Janke W., J. Chem. Phys., 2010, \textbf{133}, 184903, \doi{10.1063/1.3501368}.
\bibitem{F2} Flory P. J., Statistical Mechanics of Chain Molecules, Hanser, Munich, 1989.
\bibitem{BHM} Bosman A. W., Janssen H. M., Meijer E. W., Chem. Rev., 1999, \textbf{99}, 1665, \doi{10.1021/cr970069y}.
\bibitem{RFBH}  De Regt R.,  von Ferber C.,  Bishop M.,   Hamling T., Physica A,  2019, \textbf{516}, 50, \doi{10.1016/j.physa.2018.09.196}.
\bibitem{W1}  Wei G., Physica A, 1995, \textbf{222}, 152, \doi{10.1016/0378-4371(95)00258-8}.
\bibitem{W2}  Wei G., Physica A, 1995, \textbf{222},  155, \doi{10.1016/0378-4371(95)00259-6}.
\bibitem{BGGB}  Benhamou M.,  Ghaouar N.,  Gharbi A.,  Benmouna M., Condens. Matter Phys., 2004, \textbf{7}, 179, \\\doi{10.5488/CMP.7.1.179}.
\bibitem{WSZ}  Wawrzyriska E., Sikorski A.,  Zifferer G., Macromol. Theory Simul., 2015,  \textbf{24},  477, \\ \doi{10.1002/mats.201500036}.
\bibitem{CB} Casassa E. F., Berry G. C., J. Polym. Sci., 1966, \textbf{4, A-2}, 881, \doi{10.1002/pol.1966.160040605}.
\bibitem{FBFRZ} van Ferber  C.,  Bishop M., Forzaglia T., Reid C., Zajac G. J., Chem. Phys., 2015, \textbf{142}, 02490, \\\doi{10.1063/1.4905101}.
\bibitem{B}  Burchard W., Adv. Polym. Sci., 1983,  \textbf{48},  1, \doi{10.1007/3-540-12030-0_1}.
\bibitem{TR}Toporowski P. M.,  Roovers J., Macromolecules, 1978, \textbf{11},  365, \doi{10.1021/ma60062a017}.
\bibitem{GB}  Giupponi G., Buzza D. M. A., Macromolecules, 2002, \textbf{35}, 9799, \doi{10.1021/ma0203851}.
\bibitem{KS}Klos J. S., Sommer J. U.,  Polym. Sci. Ser. C, 2013, \textbf{55}, 125, \doi{10.1134/S1811238213070023}.
\bibitem{RA}  Rorres C.,  Anton H., Applications of Linear Algebra, John Wiley and Sons, New York, 1984.
\end{thebibliography}
\end{document}